# Anomalous T-dependence of phonon lifetimes in metallic VO$_2$

Carl Willem Rischau[1], Artem Korshunov[2], Volodymyr Multian[1], Sara A. Lopez-Paz[1], Chubin Huang[3], Lucia Varbaro[1], Jérémie Teyssier[1], Yoav Kalcheim[3], Stefano Gariglio[1], Alexei Bossak[2], Jean-Marc Triscone[1] and Javier del Valle[*4]

[1]Department of Quantum Matter Physics, University of Geneva, 24 Quai Ernest-Ansermet, 1211 Geneva, Switzerland
[2]European Synchrotron Radiation Facility, BP 220, 38043 Grenoble Cedex, France
[3]Department of Material Science and Engineering, Technion - Israel Institute of Technology, Haifa 32000, Israel
[4]Department of Physics, University of Oviedo, C/ Federico García Lorca 18, 33007 Oviedo, Spain

*Corresponding author: javier.delvalle@uniovi.es

**Abstract**

We investigate phonon lifetimes in VO$_2$ single crystals. We do so in the metallic state above the metal-insulator transition (MIT), where strong structural fluctuations are known to take place. By combining inelastic X-ray scattering and Raman spectroscopy, we track the temperature dependence of several acoustic and optical phonon modes up to 1000 K. Contrary to what is commonly observed, we find that phonon lifetimes *decrease* with *decreasing* temperature. Our results show that pre-transitional fluctuations in the metallic state give rise to strong electron-phonon scattering that onsets hundreds of degrees above the transition and increases as the MIT is approached. Notably, this effect is not limited to specific points of reciprocal space that could be associated with the structural transition.

**Introduction**

The metal-insulator transition (MIT) is among the most actively researched topics in condensed matter physics. This is mainly due to two factors: i) its potential technological applications [1–3] and ii) its complex intrinsic nature [4–8]. A paradigmatic example is that of VO$_2$, which features a well-known MIT at 340 K [4,9–11]. The MIT is accompanied by a structural transition in which chains of vanadium atoms dimerize and the lattice symmetry changes from tetragonal to monoclinic [12]. The coexisting transitions make it hard to disentangle the mechanism driving the MIT, which has resulted in a longstanding debate.

It was initially proposed that dimerization alone could account for the opening of the insulating gap [13], and that electron-lattice coupling was the main driver of the MIT, which would resemble a Peierls transition. Later calculations showed that it is necessary to include electron-electron correlations for the insulating gap to open [6,14–17], and experiments hinted that metallic VO$_2$ is a highly correlated metal on the brink of localization [18–23]. This suggested the possibility that the MIT is driven purely by electronic interactions, and the lattice just follows. Recent experiments have even suggested that MITs can take place independently of the structural transition [24–27], a scenario which has nevertheless been challenged [28–31]. This controversy between electronically-driven, or electron-phonon-driven transitions is far from being unique to VO$_2$. It is an ongoing debate for many materials featuring a MIT [7,8], but also for other systems as diverse as Kagome metals [32], charge density waves [33] or colossal magnetoresistance manganites [34].

In its high temperature tetragonal phase, VO$_2$ is a bad metal with resistivity above the Mott-Ioffe-Regel limit and a broad Drude peak [19–21,23]. Its structure has been shown to feature strong fluctuations [5,35–37], which can be interpreted as transient excursions into the low temperature monoclinic structure [35]. They manifest as intense diffuse (111) scattering planes passing through the R and M points of the tetragonal Brillouin zone [35] (See Figure 1a). Budai *et al*. experimentally showed that these fluctuations arise from flat and highly anharmonic acoustic phonon branches, inherent to the rutile tetragonal structure of metallic VO$_2$ [36]. Such disorder and anharmonicity leave an imprint in the vibrational spectrum: phonon lifetimes in the metallic state are extremely short, to the point of being comparable to the oscillation period. Raman spectra of metallic VO$_2$ look almost featureless due to the broadened peaks [29,38].

A proper understanding of pre-transitional fluctuations is key for solving the MIT puzzle. A detailed characterization of the temperature ($T$) evolution of phonon dynamics is crucial to this end. Despite this, few works have investigated the vibrational properties of metallic VO$_2$, which are usually characterized only in the immediate vicinity of the MIT [38]. Within a few tens of degrees around the MIT, the temperature evolution of phonon dynamics seems to not follow any trend. However, this may be expected given the strong first order, discontinuous character of the VO$_2$ MIT. Measuring over a larger $T$ range, deeper into the metallic state, might unveil new information.

Here we use a combination of inelastic X-ray scattering (IXS) and optical Raman spectroscopy to track the lifetime of several acoustic and optical phonons from the MIT up to 973 K. We find that upon increasing temperature, phonon linewidths not only do not increase -as would be expected-, but in many cases decrease. This is observed for multiple optical and acoustic modes. Since phonon-phonon scattering rates decrease monotonically with lowering temperature, we conclude that electron-phonon scattering increases as the MIT is approached. This takes place over several hundreds of degrees and is observed in several unrelated points of reciprocal space. Our results reveal an intimate relationship between electron-phonon scattering and the MIT and set the temperature scale for pre-transitional fluctuations in metallic VO$_2$, underlining the complex nature of this phase.

**Experiments**

We synthesized VO$_2$ single crystals using the self-flux method [39,40], by annealing V$_2$O$_5$ powder (99.95 %, Aldrich) for 24 hours at 1000º C under an Argon gas flow. Large, high quality single crystals with lengths up to several mm can be obtained this way (Figure 1b). Figure 1c shows resistivity vs temperature of one of them, featuring a sharp, five orders of magnitude resistivity drop at the MIT.

We first focus on the temperature dependence of three acoustic phonons. We chose the R, M and A high symmetry points of the tetragonal Brillouin zone (Figure 1a). These modes belong to the flat phonon bands reported by Budai *et al* [36]. The M and R points are part of the (111) diffuse scattering planes, so they are candidates to exhibit signs of criticality. Since these three points have non-zero momentum, high energy photons are required to probe them. We performed IXS measurements at the ID28 beamline at ESRF. We shone a 17.8 keV beam on the crystals and positioned the detector arm to receive scattered X-rays coming from the R (0,2.5,1.5), M (3.5,0.5,0) or A (2.5,0.5,0.5) points. We used a crystal analyzer and temperature-controlled monochromator to finely scan the photon energy at constant momentum, with a resolution of 3 meV. A $N_2$ heatblower was used to warm up and control the sample temperature while preventing its oxidation. None of the crystals showed signs of degradation after being warmed to 973 K.

Several scans are shown in Figures 2a (R point) and 2b (M point). Two temperatures are shown in each case: 373 K (100º C) and 773 K (500º C). The symmetric peaks observed in the scans correspond to the processes of phonon absorption (Stokes) and emission (Anti-Stokes). Comparing scans at 373 K and 773 K, a slight phonon softening can be appreciated for lower temperatures, in accordance to previously reported data [36]. On a first inspection, the overall width of the peaks does not seem to change with temperature, which is surprising given the large temperature difference. Figure 2c focuses on the right shoulder of the Stokes peaks. The horizontal axis has been shifted to center the peaks around E=0 meV, making it easier to compare their widths at different temperatures. While at high temperatures the R point peak experiences a slight widening, a clear thinning is observed at the M point. The continuous lines are fits using a single-damped-harmonic-oscillator model convoluted with the experimental resolution. From them we get that for the R point, the full width at half maximum (FWHM) is 5.2 meV at 373 K and 6.1 meV at 773 K. For the M point, it goes from 4.9 meV at 373 K down to 3.5 meV at 773 K.

Figure 3 shows scans at the A point, also for *T*=373 K and 773 K. In this case, at least two partially overlapping phonons are present. We must note that only the E>0 region was scanned at this reciprocal space location. Similarly to what happens at the M point, peaks are wider at 373 K than at 773 K. This can be better seen in the inset of Figure 3, where the peak position was offset in the horizontal axis.

Optical phonon lifetimes were investigated using Raman spectroscopy, which probes phonons at the Γ point. Raman spectra were recorded using a Horiba LabRAM HR Evolution spectrometer with an excitation wavelength of 532 nm. A Linkam TS1500 optical stage was used to reach high temperatures. Figures 4a and 4b show Raman spectra with light polarized along the (001) and (100) directions, respectively. Multiple temperatures between 373 K and 873 K are shown. Peaks are broad and their intensity is low compared to the background. Nonetheless, it is possible to distinguish the four Raman active modes allowed in the rutile structure. They are located at around 27 ($B_{1g}$), 44 ($E_g$), 59 ($A_{1g}$) and 78 meV ($B_{2g}$). The $A_{1g}$ mode is strongly suppressed for light polarized along (001), but dominates the spectrum along (100), showing a rather asymmetric profile, not unusual for strongly anharmonic phonons [41,42]. This, together with the peak broadness, prevents us from fitting the (100) polarized spectra reliably. However, good fitting is possible along the (001) direction. One such fits, obtained using the Reffit software [43], is shown in Figure 4a for 773 K as example.

Comparing spectra at different temperatures, we find a similar behavior as in IXS data. Upon increasing temperature, Raman peaks do not broaden, as usually observed. They either remain similar or they sharpen. The latter can be seen for spectra in Figure 4a, where peaks at 27, 44 and 78 meV are noticeably better resolved at 773 K than at 373 K. As an example, the FWHM of the 78 meV peak, as obtained from fitting, goes down from 27 meV at 773 K to 22 meV at 373 K.

**Discussion**

Figure 5a shows phonon lifetimes as a function of temperature, for six different optical and acoustic phonons. We defined the lifetime as $\tau = \hbar/$FWHM, where $\hbar$ is the Planck constant. Figure 5b shows the peak positions vs $T$. Peak positions and FWHM were calculated by fitting. Since Raman spectra show multiple overlapping peaks, $\tau$ are estimated with much higher error compared to IXS data. The A point acoustic phonon is not included in this figure since the multiple overlapping peaks made it hard to reliably determine $\tau$, but it qualitatively follows the same trend as the M point.

Four of the phonons -five if we also consider the A point- have a similar trend: $\tau$ slowly decreases as temperature is lowered. For the $A_{1g}$ phonon it stays constant with temperature, while for the R point acoustic phonon it slightly increases, although plateauing near the MIT. At high enough temperatures (>800-900 K) a more conventional $\tau(T)$ trend seems to be partially recovered. To put this anomalous behavior into context, the well-known Silicon Raman mode at 530 cm$^{-1}$ almost doubles its lifetime when lowering $T$ from 800 K to 100 K [44].

Neglecting impurity and boundary effects, the total phonon scattering rate is given by the addition of the phonon-phonon and electron-phonon scattering rates [45]. Phonon-phonon scattering increases with phonon population, which is given by $n = \left(e^{E_{Ph}/kT} - 1\right)^{-1}$, where $E_{ph}$ is the phonon energy and $k$ is the Boltzmann constant. This implies that phonon-phonon scattering must decrease monotonically as the temperature is lowered. Our experimental results show that total phonon scattering rates slowly increase as $T$ is lowered. Therefore, we conclude that electron-phonon must increase markedly as the MIT is approached, as schematically depicted in Figure 5c.

Similar behaviors have been observed in systems featuring charge density waves, where nesting and electron-phonon coupling is known to be crucial for the transition [33,46–49]. However, the anomalous $\tau$ dependence in these systems is usually accompanied by significant phonon softening and is limited to certain points in reciprocal space associated with the structural transition. In our case one could expect this effect to be stronger around the R point, since it lies in the flat acoustic branches and it is related to the static distortions of the insulating monoclinic phase [36], but that is not the case. Surprisingly, we observe the anomalous behavior for multiple, unrelated points of reciprocal space, and regardless of whether the phonon softens or hardens.

The reason why this effect is more pronounced for some phonons than for others remains unclear. It is likely to depend on anharmonic couplings between them. If a specific mode is coupled to many others, it may be especially susceptible to phonon-phonon scattering. It is possible that for such mode, phonon-phonon scattering increases faster with temperature, compared to other modes. In turn, this could counteract the temperature trend of electron-phonon scattering and make the τ(*T*) curve take a more conventional shape. A detailed understanding of these couplings might reveal important clues about pre-transitional fluctuations in metallic $VO_2$, but it is beyond the scope of this experimental paper.

We must note that an increasing electron-phonon scattering upon approaching the transition does not necessarily mean that the purely electronic-driven MIT scenario can be discarded. Let´s consider that pure electron-electron interactions were driving the MIT. In such case, the metallic state could be expected to feature pure charge fluctuations with a symmetry different than tetragonal. Phonons would see this change in the charge density as a defect, since it doesn't follow the lattice symmetry, and they would therefore scatter off it. As this charge redistribution increases on cooling when approaching the transition, so would the scattering rate.

**Conclusions**

We have performed inelastic X-ray scattering and Raman spectroscopy to track the temperature dependence of several acoustic and optical phonons in metallic $VO_2$. We observe an anomalous behavior: phonon lifetimes decrease as the temperature is lowered towards the transition. Our results show that, within the metallic state, electron-phonon scattering increases as the MIT is approached. The effect is observed up to very high temperatures, which suggests that strong pre-transitional fluctuations onset at least 500 K above the transition, without signs of divergence or criticality close to the MIT. Remarkably, this phenomenon is not limited to specific areas of the reciprocal space or to phonons that soften or are related to the structural transition. This is potentially due to the strong anharmonic coupling between different vibrational modes. Our work underlines the close relationship between electron-phonon coupling and the MIT and unveils the temperature scale in which pre-transitional fluctuations set in. These are key elements for constructing a complete model capturing the physics of the MIT.

**Acknowledgements**

We thank Lukas Korosec, Giacomo Mazza and Marios Hadjimichael for insightful discussions. We thank Bertina Fisher and Fabian von Rohr for support during the sample synthesis. This work was funded by the Spanish Ministry of Science through a Ramón y Cajal Fellowship (Grant No. RYC2021-030952-I), by the Swiss National Science Foundation through an Ambizione Fellowship (#PZ00P2_185848), and by the Asturias FICYT under Grant No. AYUD/2021/51185 with the support of FEDER funds. C.W.R. was supported by the U.S. Office of Naval Research through the NICOP Grant N62909-21-1-2028. We thank the European Synchrotron Radiation Facility (ESRF) for provision of synchrotron radiation facilities under proposal number HC-5354. Crystal growth at Technion received funding from the European Union's Horizon Europe Research and Innovation Program under grant agreement no. 2031928. L.V. was supported by the Swiss National Science Foundation - division II (200020_179155 and 200020_207338).

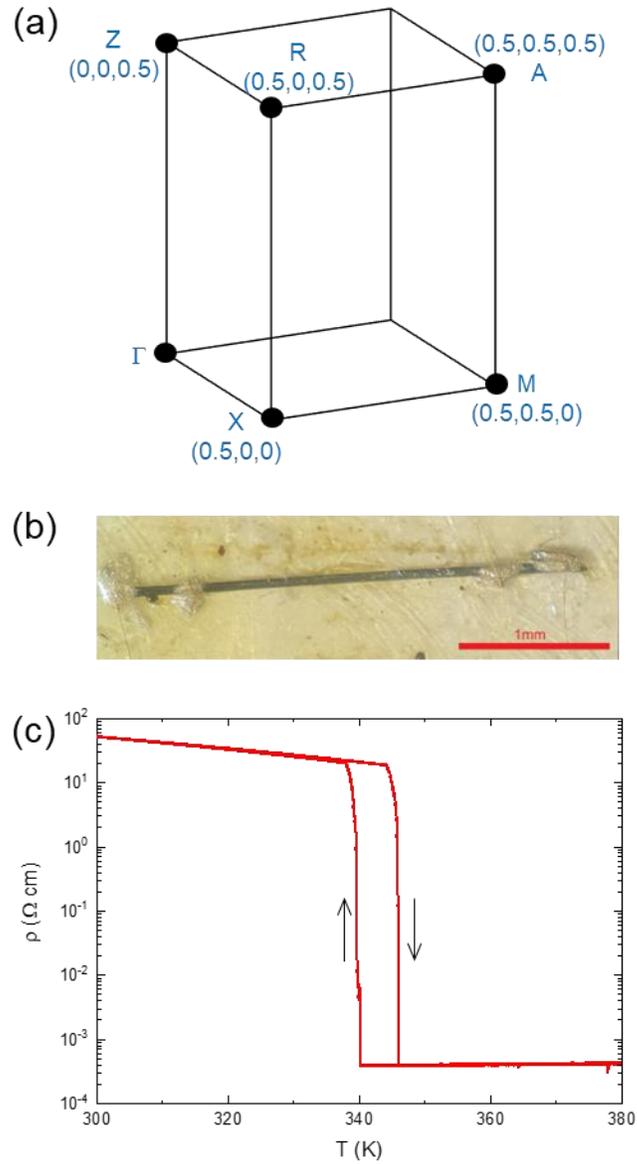

FIG. 1. (a) Schematic representation of the high symmetry points within the first Brillouin zone of tetragonal metallic VO$_2$. (b) Optical microscope image of a VO$_2$ single crystal, with four silver paste contacts for resistivity measurements. The red scale bar corresponds to 1 mm. (c) Resistivity $\rho$ vs temperature. Arrows indicate the scan direction on cooling and warming.

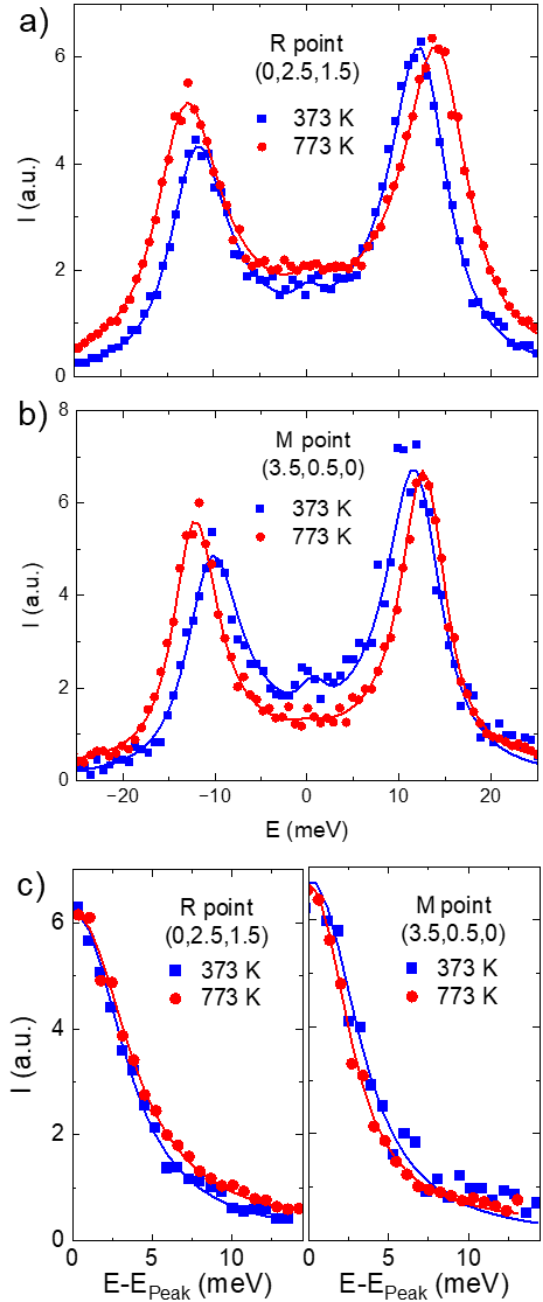

FIG. 2. IXS data at the R and M points. (a) and (b) are constant momentum energy scans at the (0,2.5,1.5) R point and the (3.5,0.5,0) M point, respectively. Scans at 373 K (100°C) and 773 K (500°C) are shown for both. Intensities are rescaled to easily compare different temperatures. Continuous lines are fits using a single-damped-harmonic-oscillator model convoluted with the experimental resolution. An elastic line at E=0 was included in the fit. (c) Energy scans of the right shoulder of the Stokes peak. The horizontal axis was shifted by the peak energy (E-$E_{Peak}$ (T)), so that peak widths at 373 K and 773 K could be easily compared.

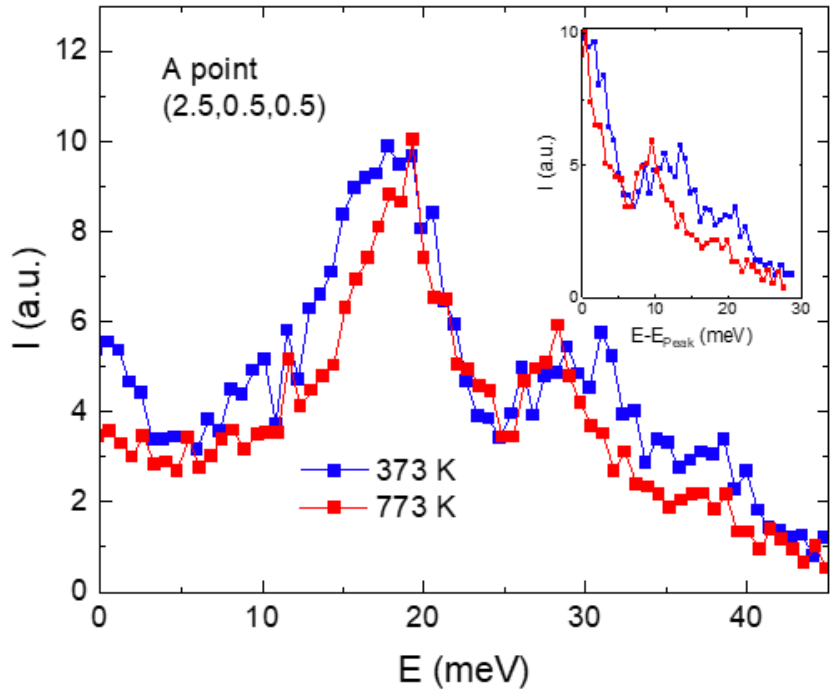

FIG. 3. IXS data at the A point. Constant momentum energy scans at the (2.5,0.5,0.5) A point. Scans at 373 K and 773 K are shown. Intensities are rescaled to easily compare the two temperatures. Inset: energy scans of the right shoulder of the Stokes peak. The horizontal axis was shifted by the peak energy (E-$E_{Peak}$ (T)), so that peak widths at 373 K and 773 K could be easily compared.

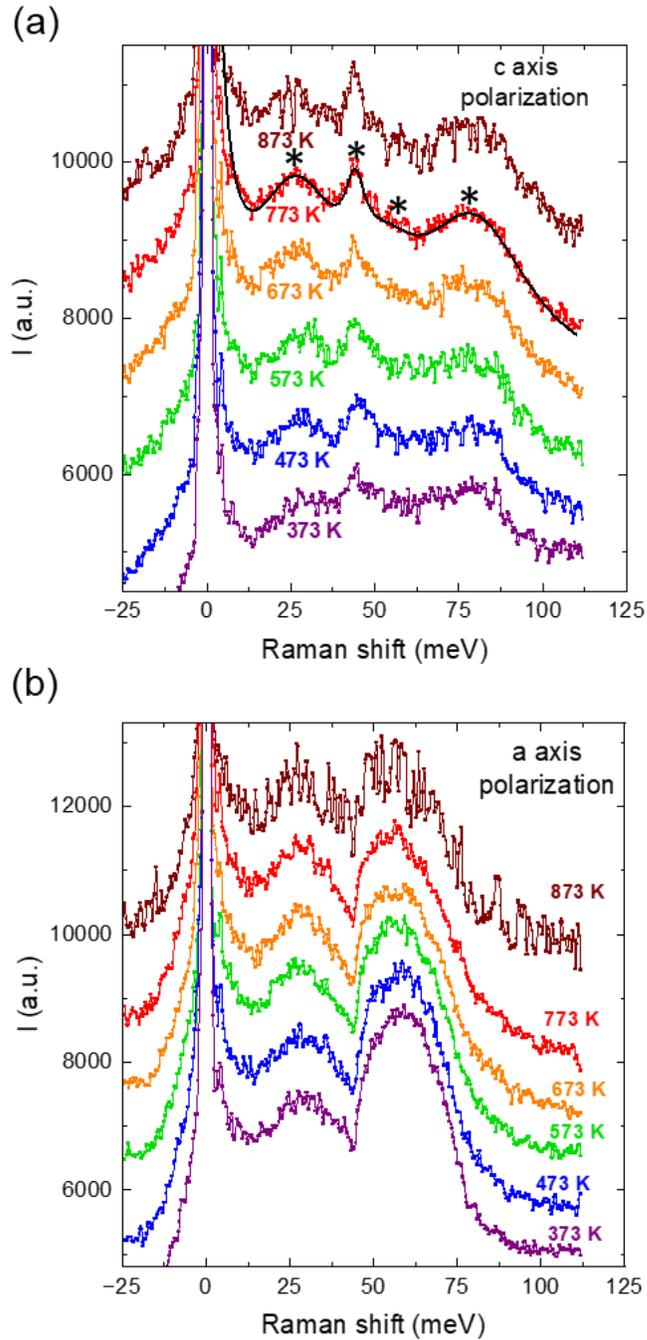

FIG. 4. Raman spectra taken with light polarized along the c axis (a) and a axis (b). For each polarization, multiple spectra between 373 K and 873 K are shown. Spectra have been vertically shifted for clarity. The black continuous line following the 773 K spectrum of panel (a) is an example of fit done with the Reffit software. The asterisks in panel (a) show where the four active modes are located.

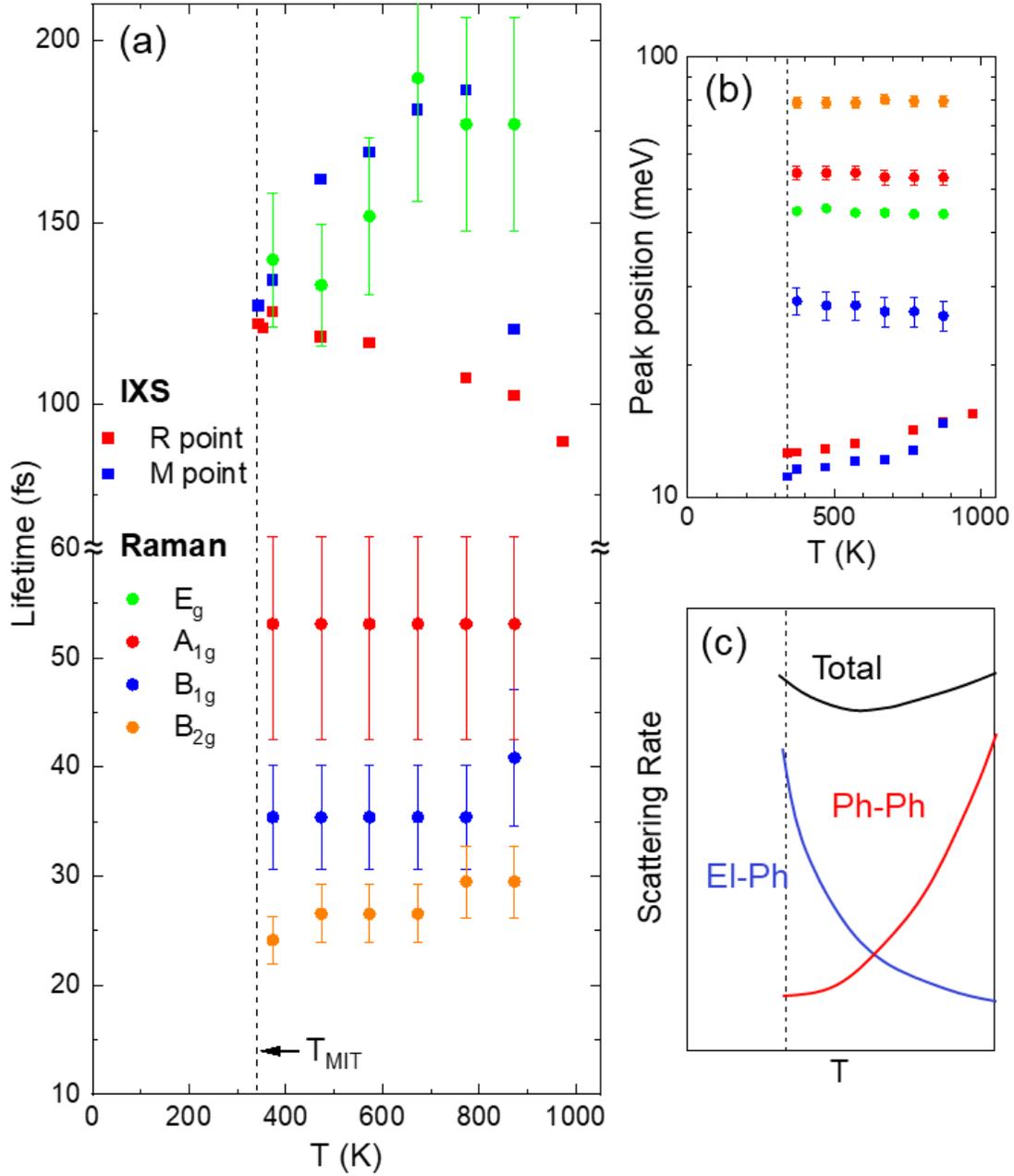

FIG. 5. (a) Phonon lifetimes vs temperature for six phonon modes. The vertical dashed line indicates the MIT temperature. Since Raman spectra show multiple overlapping peaks, lifetimes ($\tau$) are estimated with much higher error compared to IXS data. Note that there is a change in the y axis scale above 60 fs. (b) Peak position vs temperature for the same modes. Color legend is the same as in panel (a). (c) Schematic representation of the temperature dependence of phonon-phonon (red), electron-phonon (blue) and total scattering rates (black).